\documentclass[twocolumn, aps, float, superscriptaddress]{revtex4}

\usepackage{mathptmx}
\usepackage{subfigure}
\usepackage{dcolumn}
\usepackage{amsmath,amssymb}
\usepackage{bm}
\usepackage{color}
\usepackage{latexsym}
\usepackage[english]{babel}
\usepackage{times}

\usepackage{psfrag,graphicx}
\usepackage{epsf}
\usepackage{amsfonts}
\usepackage{natbib}
\usepackage{epstopdf}
\usepackage{appendix}
\usepackage[colorlinks=true,citecolor=blue,linkcolor=blue]{hyperref}

\begin{document}

\pacs{05.30.Jp, 03.75.Mn, 67.85.-d, 71.70.Ej}

\title{The Raman dressed spin-1 spin-orbit coupled quantum gas}

\author{Zhihao Lan}
\affiliation{Mathematical Sciences, University of Southampton, Highfield, Southampton, SO17 1BJ, United Kingdom}
\email{z.lan@soton.ac.uk, lanzhihao7@gmail.com}
\author{Patrik \"Ohberg}
\affiliation{SUPA, Institute of Photonics and Quantum Sciences, Heriot-Watt University, Edinburgh EH14 4AS, United Kingdom}

\begin{abstract}

The recently realized spin-orbit coupled quantum gases (Y.-J Lin {\it et al}.,  Nature 471, 83-86 (2011); P. Wang {\it et al}.,  PRL 109, 095301 (2012); L. W. Cheuk {\it et al}.,  PRL 109, 095302 (2012)) mark a breakthrough in the cold atom community. In these experiments, two hyperfine states are selected from a hyperfine manifold to mimic a pseudospin-1/2 spin-orbit coupled system by the method of Raman dressing, which is applicable to both bosonic and fermionic gases. In this work, we show that the method used in these experiments can be generalized to create any large pseudospin spin-orbit coupled gas if more hyperfine states are coupled equally by the Raman lasers.  As an example,  we study in detail a quantum gas with three hyperfine states coupled by the Raman lasers, and show when the state-dependent energy shifts of the three states are comparable, triple-degenerate minima will appear at the bottom of the band dispersions, thus realizing a spin-1 spin-orbit coupled quantum gas. A novel feature of this three minima regime is that there can be two different kinds of stripe phases with different wavelengths, which has an interesting connection to the ferromagnetic and polar phases of spin-1 spinor Bose-Einstein condensates without spin-orbit coupling.

\end{abstract}

\maketitle
\section{Introduction}
 Cold atoms have proven to be ideal platforms to  simulate a variety of phenomena ranging from condensed matter  to nuclear physics due to their unprecedented controllability~\cite{maciej_ap, bloch_np}.
Among these, a great amount of theoretical and experimental efforts in recent years have been dedicated to the engineering of gauge potentials for neutral atoms ~\cite{gauge_rmp, review_exp}. Not only because it provides a platform to simulate magnetic fields and spin-orbit couplings (a special form of non-Abelian gauge potentials), and related phenomena such as quantum spin Hall effects or topological phases~\cite{TI, nathan, she} in condensed matter physics, but also due to the dramatic impact the gauge potentials have on the system dynamics. For instance,  spin-orbit effects with bosons have no counterpart in the electronic properties of solids. Even at single particle level the introduction of gauge potentials  will modify the particle dispersions, leading to exotic properties such as negative reflection~\cite{negative_r}, or multi-refringence~\cite{high_spin, kennett}. These modified particle dispersions will have dramatic effects on the few-body or many-body physics when interactions are present. Indeed, the enhanced density of states by the gauge potentials leads to two-body bound states even at the BCS (Bardeen-Cooper-Schrieffer) side of the resonance \cite{bs_shenoy}. Moreover, the broken Galilean invariance due to the presence of  the gauge potentials could result in finite-momentum Cooper pairs \cite{fm_pu}. As a consequence, the possibility of BEC (Bose-Einstein-Condensates) to BCS crossover by increasing the strength of the gauge coupling \cite{crossover_chuanwei, crossover_pu, crossover_zhai}, or the possible realization of the long-sought FFLO (Fulde-Ferrell-Larkin-Ovchinnikov) superfluidity \cite{FFLO_chuanwei, FFLO_yi}  at many-body level have attracted great interests.

To date, synthetic spin-orbit couplings have been realized experimentally for both BEC and Degenerate Fermi Gases (DFG)~\cite{review_exp}.  The key idea behind these achievements, i.e.,  Raman dressing,  was first demonstrated in a  series of experiments  by Lin et al. ~\cite{NIST_prl, NIST_n1, NIST_np, NIST_n2} for a $^{87} $Rb BEC and  later on for fermionic $^{40}$K by \cite{ShanXi_fermi} and with $^{6}$Li by \cite{MIT_fermi}. The elegance of this method is in its simplicity where only one pair of lasers and an external magnetic field are used. Both the Abelian and Non-Abelian regimes can be reached by tuning the laser power. 
For example, in the first three experiments at NIST, only a single minimum of the lowest energy dispersion in the form of $\hbar^2(k_x-A_x)^2/2m$ was created, where  a constant $A_x$, space-dependent $A_x$ and time-dependent $A_x$, lead to, a uniform vector potential with zero magnetic field, non-zero magnetic field $B_z=-\partial_y A_x\neq 0$, and non-zero electric field $E_x=-\partial_t A_x\neq 0$ respectively. Conversely, two minima in the energy dispersion is interpreted as two dressed spin states responsible for the synthetic spin-orbit coupling with equal Rashba and Dresselhaus strengths. We note that the most recent experimental ~\cite{USTC_boson, quench_so} and theoretical~\cite{ho,sinha2011,li, zhai_two} studies using this Raman scheme are mostly concerned with the two minima regime. It is, however, not an insurmountable task to experimentally control all the three Zeeman levels \cite{ian_pc}, and by doing so obtain a spin-1 scenario. Magnetically generated spin-orbit coupling \cite{anderson2013} may also provide a viable route to larger spin systems. 

In this work, we show that a three minima regime in the energy dispersion of the NIST setup can be reached. We show in detail how this three minima regime can emerge as a function of the Raman strength $\Omega_R$ and the quadratic Zeeman energy $\epsilon$ when the detuning $\delta$ is zero and the contributions of the three Zeeman states of the underlying manifold are comparable. This is in contrast to the extensively studied phase diagram  in the $\Omega_R$-$\delta$ plane \cite{NIST_n2, USTC_boson, quench_so}. For a special configuration of the parameters,  a triple-degenerate minima can appear at the bottom of the spectrum, thus realizing a spin-1 spin-orbit coupled quantum gas.  Our work shows that the method of Raman dressing can readily be used to synthesize large pseudospin-orbit couplings for neutral atoms if more hyperfine states are coupled equally.
\begin{figure}[!hbp]
\centering
\includegraphics[width=1\columnwidth]{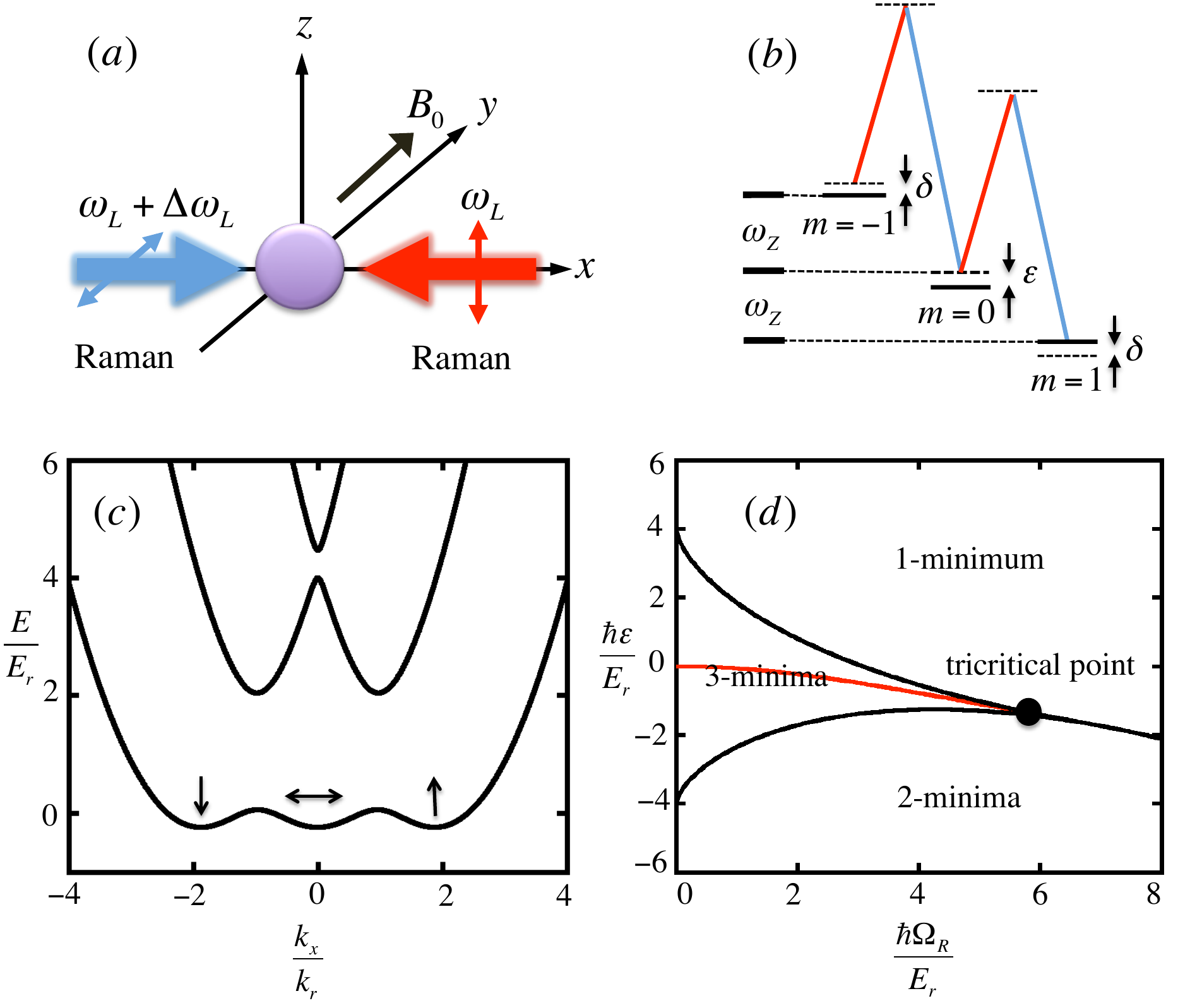}
\caption{ (color online) (a) Schematic of the experimental setup at NIST.  Two counter-propagating Raman lasers with frequencies $\omega_L$ and $\omega_L+\Delta \omega_L$ along $\hat{x}$ impinge on the atomic cloud. A bias field $B_0$ along  $\hat{y}$ produces the Zeeman effects. (b) Level diagram of the Raman coupling scheme within the F=1 manifold.   The linear and quadratic Zeeman shifts are  $\omega_Z$ and $\epsilon$ while  the detuning from Raman resonance  $\delta$,  which is set to zero in this study. Atoms excited by the Raman lasers will change their spin projection along the magnetic field by 1 while increase their linear momentum by $2\hbar k_r$.  (c) The spectrum of $\mathcal{H}(\tilde{k}_x)$ with $\hbar\Omega_R=2E_r $ and $\hbar\epsilon=-0.23E_r$. The triple-degenerate minima at the bottom of the spectrum serves as a spin-1 system. (d) The single particle phase diagram in the plane of $\Omega_R$-$\epsilon$. The three phases meet at a tricritical point  and the red line shows the regime where the three minima are degenerate in energy.}
\label{regime}
\end{figure}

\section{The three-minima regime}
We follow the NIST setup shown in Fig.~\ref{regime} (a), where two counter-propagating Raman lasers along $\hat{x}$ with frequency difference $\Delta\omega_L$ and momentum difference $2\hbar k_r$ couple the three hyperfine states of a ultracold atomic cloud. At single particle level, the setup is applicable to both bosonic and fermionic gases as long as suitable hyperfine states can be selected. For simplicity, we will denote the three states as an $F=1$ manifold, $|F, m_F \rangle =|1,-1 \rangle, |1, 0 \rangle, |1,+1 \rangle$, though in principle they can be three hyperfine states of a much higher manifold. Meanwhile there is a magnetic field along $\hat{y}$ producing the linear and quadratic Zeeman shifts $\hbar\omega_Z$ and $\hbar\epsilon$ for the three states.  The lasers induce a Raman transition in the atom, transferring linear momentum $2\hbar k_r\hat{x}$ to the atom while increasing its spin angular momentum by $\hbar$ at the same time ( Fig.~\ref{regime} (b)). The Hamiltonian of the system dressed by the Raman lasers is described by:  $\mathcal{H}=\hbar^2k^2/2m-\hbar\Omega_R/2[e^{i(2q_rx-\Delta\omega_L t)}(\mathcal{F}_z+i\mathcal{F}_x)+H.c.]-\hbar\omega_Z \mathcal{F}_y+\hbar\epsilon \mathcal{F}_y^2$, where $\mathcal{F}$ is the spin-1 operator, $\Omega_R$ the Raman frequency associated with the Raman process, $\hbar\omega_Z$ and $\hbar\epsilon$  the linear and quadratic Zeeman energies of the three levels. In the rotating wave approximation for the frame in spin space rotating about $\hat{y}$ with frequency $\Delta\omega_L$, the Hamiltonian becomes static, $\mathcal{H}=\hbar^2k^2/2m-\hbar\Omega_R[\cos(2q_rx)\mathcal{F}_z-\sin(2q_rx)\mathcal{F}_x]-\hbar(\omega_Z-\Delta\omega_L) \mathcal{F}_y+\hbar\epsilon \mathcal{F}_y^2$.  We can then apply another rotation in spin space along $\hat{y}$ with angle $2\hbar k_r$, and the Hamiltonian in the state basis of $\Psi(\tilde{k}_x)=\{ |-1, \tilde{k}_x-2k_r\rangle, |0, \tilde{k}_x\rangle, |1, \tilde{k}_x+2k_r\rangle \}$, labeled by the wave vector of quasimomentum $\tilde{k}_x$, reduces to $\mathcal{H}=\mathcal{H}(\tilde{k}_x)+[\hbar^2(k_y^2+k_z^2)/2m] $,  with  $\mathcal{H}(\tilde{k}_x)$ for the Raman coupling given by \cite{NIST_prl},

\begin{equation}
\mathcal{H}(\tilde{k}_x)=\hbar
\begin{pmatrix}
\frac{\hbar(\tilde{k}_x+2k_r)^2}{2m}-\delta & \Omega_R/2 & 0 \\
\Omega_R/2 & \frac{\hbar \tilde{k}_x^2}{2m}-\hbar\epsilon & \Omega_R/2\\
0 & \Omega_R/2 & \frac{\hbar(\tilde{k}_x-2k_r)^2}{2m}+\delta
\label{hamiltonian}
\end{pmatrix}
\end{equation}
where  $\delta=(\Delta\omega_L-\omega_Z)$ is the detuning from Raman resonance. It is to be noted that the parameters $\delta$ and $\epsilon$ just shift the three bare branches up and down.  While the  experiments~\cite {USTC_boson, quench_so, NIST_n2} use $\delta$ to select two out of the three Zeeman states as a spin-1/2 system,  here we set $\delta=0$ and leave only $\epsilon$ as a free parameter to balance the contributions of the three bare branches, which  in principle can be controlled by state-dependent trapping potentials, i.e., both positive and negative $\epsilon$ can be realized in this way. Alternatively,  negative quadratic Zeeman energy can also be  realized experimentally by the technique of microwave dressing (e.g., see~\cite{microwave_dressing}).  In the following, we define $E_r=\hbar^2k_r^2/2m$ as the recoil energy which will be used as the energy scale and  $k_r$ the momentum scale.

Fig.~\ref{regime} (d) shows the single particle phase diagram of Hamiltonian~(\ref{hamiltonian}) in the plane of Raman coupling ($\Omega_R$) and energy shift of the middle branch ($\epsilon$).  The three phases,  characterized by one minimum, two minima and three minima  in the lowest energy dispersion, meet at a tricritical point beyond which the three minima regime no long exists. The red line in Fig.~\ref{regime}  (d) shows the regime where the three minima are exactly degenerate  in energy. We show in Fig.~\ref{regime}  (c) one example of the triple degenerate minima regime with parameters of $\hbar\Omega_R=2E_r$ and $\hbar\epsilon=-0.23E_r$. In this case, the three degenerate minima at the bottom of the spectrum serve as a spin-1 manifold and the atomic gas is spin-orbit coupled with enlarged pseudospin of 1.   We note our phase diagram in the plane of $\Omega_R$-$\epsilon$ is very different from the phase diagram in the plane of  $\Omega_R$-$\delta$ (e.g., see ~\cite{USTC_boson}). The phase diagram in the plane of $\Omega_R$-$\delta$ shows only two phases, a two local minima regime or a one minimum regime and  when the Raman coupling is strong enough, only one single minimum can exist.  We also note tricriticality and similar phase diagrams in spin-orbit coupled BEC are  discussed recently by Li {\it et al} ~\cite{li}, but in a two minima regime at many particle level, which is different from our three minima regime at the single particle level.

The physical reason of the phase diagram in Fig.~\ref{regime}  (d) can be understood as follows.  At $(\Omega_R, \epsilon)=(0,0)$, the energy dispersions are the three bare parabolas located at $\tilde{k}_x^{min}/ k_r=-2, 0, 2$. With increasing $\Omega_R$, gaps will open at the anti-crossing points and $\epsilon$ will shift the middle branch up with negative $\epsilon$ or down with positive $\epsilon$.  When the middle branch shifts up (with a decreasing $\epsilon$) the minimum located at $\tilde{k}_x=0$ will merge with the two neighbouring maxima into a single maximum, thus the system enters the two minima regime. Conversely, when the middle branch shifts down (with increasing $\epsilon$) the two minima located at $\tilde{k}_x^{min}/ k_r=-2, 2 $ will merge with its neighboring maximum and leave only a minimum at $\tilde{k}_x=0$, thus the system enters the single minimum regime. When the Raman coupling is strong enough, the minimum located at $\tilde{k}_x=0$ will be destroyed by the anti-crossing between the dispersion curves, and as a result, the three minima regime no longer exists.

\begin{figure}[!hbp]
\centering
\includegraphics[width=1\columnwidth]{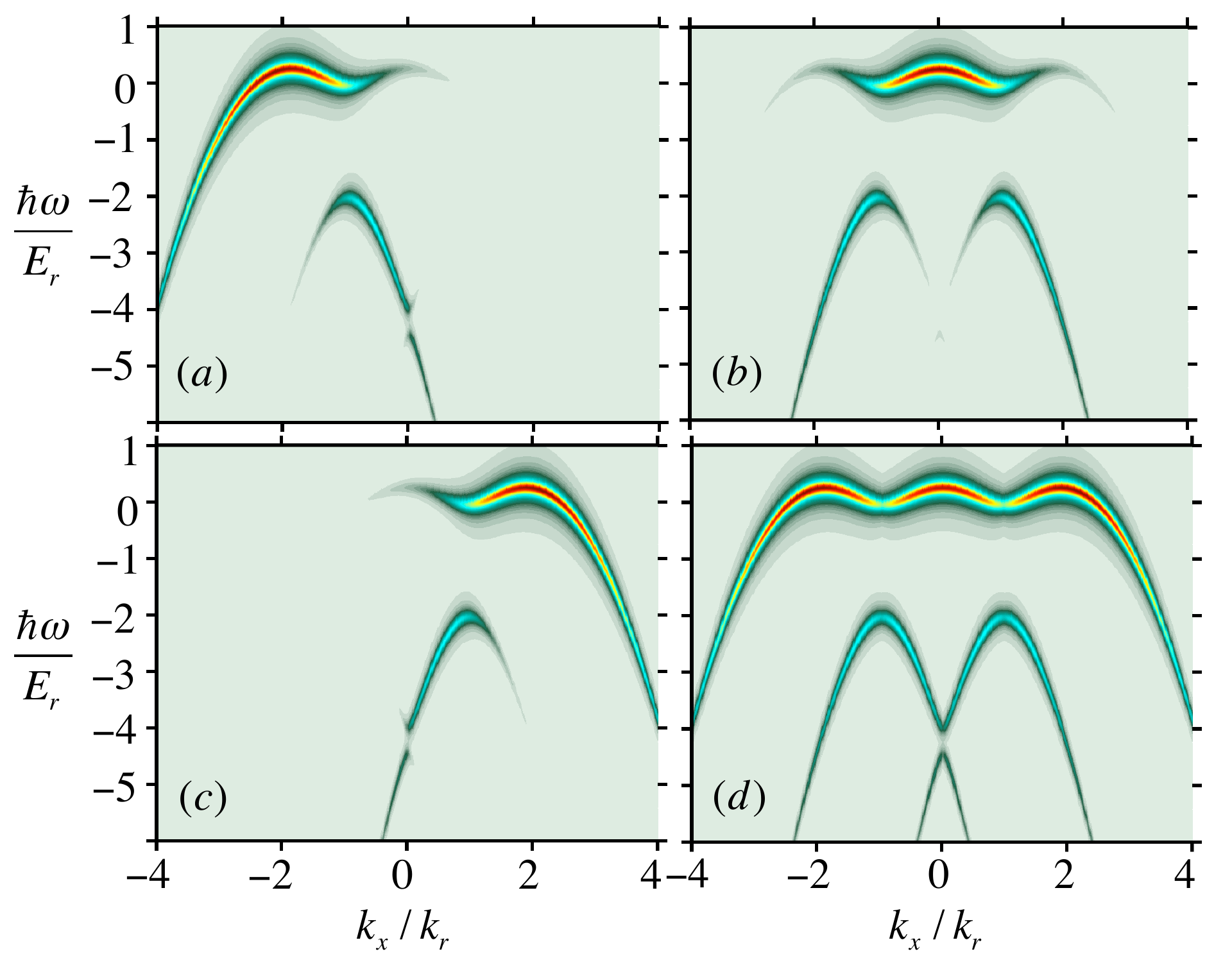}
\caption{ (color online) Momentum-resolved radio-frequency (rf)  spectroscopy for reconstructing the band dispersions in Fig.~\ref{regime}(c) for a spin-orbit coupled Fermi gas.  Plots (a), (b) and (c) show momentum-resolved rf spectroscopies for the three Zeeman states respectively, while (d) shows the reconstructed band dispersions by the combination of (a), (b) and (c). Parameters used: $\hbar\Omega_R=2E_r $, $\hbar\epsilon=-0.23E_r$, chemical potential $\mu=3E_r$ and temperature $T=0.6\mu$ and in consideration of the energy resolution of the spectroscopy $\gamma \sim 0.1E_r$. We have replaced the $\delta$ function for the energy conservation by $\delta(x)=(\gamma/\pi)/[x^2+\gamma^2]$~\cite{MIT_fermi, liu_rf}. Note when increasing the chemical potential, the transfer strength will become stronger and the higher branches will also get occupied, since there will be more and more atoms in the system. }
\label{rf}
\end{figure}

\section{Momentum-resolved radio-frequency (rf)  spectroscopy}
By using the same method of Raman dressing as in the NIST experiments, spin-orbit coupled Fermi gases have also been realized experimentally at ShanXi University ($^{40} $K ) ~\cite{ShanXi_fermi} and Massachusetts Institute of Technology (MIT) ($^6$Li)~\cite{ MIT_fermi} where the  band dispersions have been studied by momentum-resolved rf spectroscopy and spin injection spectroscopy respectively. While the spin-injection spectroscopy uses a rf  laser  to inject free atoms in a reservoir state into the empty spin-orbit coupled system, after which the momentum and spin of injected atoms are mapped out by using time of flight and spin-resolved detection, the momentum-resolved rf spectroscopy uses a rf  laser to transfer atoms from one of the hyperfine states for constructing the spin-orbit coupling system to an empty reservoir state. For a non-interacting system, the momentum-resolved rf spectroscopy yields equivalent information to the spin-injection spectroscopy. 

We have calculated the momentum-resolved rf  spectroscopy in order to show experimentally it is possible to observe the three-minima band structure shown in Fig.~\ref{regime}(c) (see~\cite{liu_rf} for details of the calculation, and \cite{jing_rf} for recent experiment).  We present the results in Fig.~\ref{rf}, where plots (a), (b) and (c) show the momentum-resolved rf spectroscopy for the three Zeeman states that are selected to synthesize the spin-orbit coupling  and  (d) shows the reconstructed band structure by the combination of plots  (a), (b) and (c). The reason for reconstructing the band dispersions in Fig.~\ref{regime}(c) by using the rf spectroscopies of  all the three Zeeman states is because each branch of the band dispersions is a mixture of the bare dispersions of the three Zeeman states.  From Fig.~\ref{rf}(d), we see the qualitative features of the band structure in Fig.~\ref{regime}(c)  are clearly visible. The rf spectrum also shows an important feature of the system, i.e., the weights of the three dressed spin states are largely determined by the three bare branches, which is the essential point  for the emerging of two different stripe phases to be discussed later.

\begin{figure}[!hbp]
\centering
\includegraphics[width=1\columnwidth]{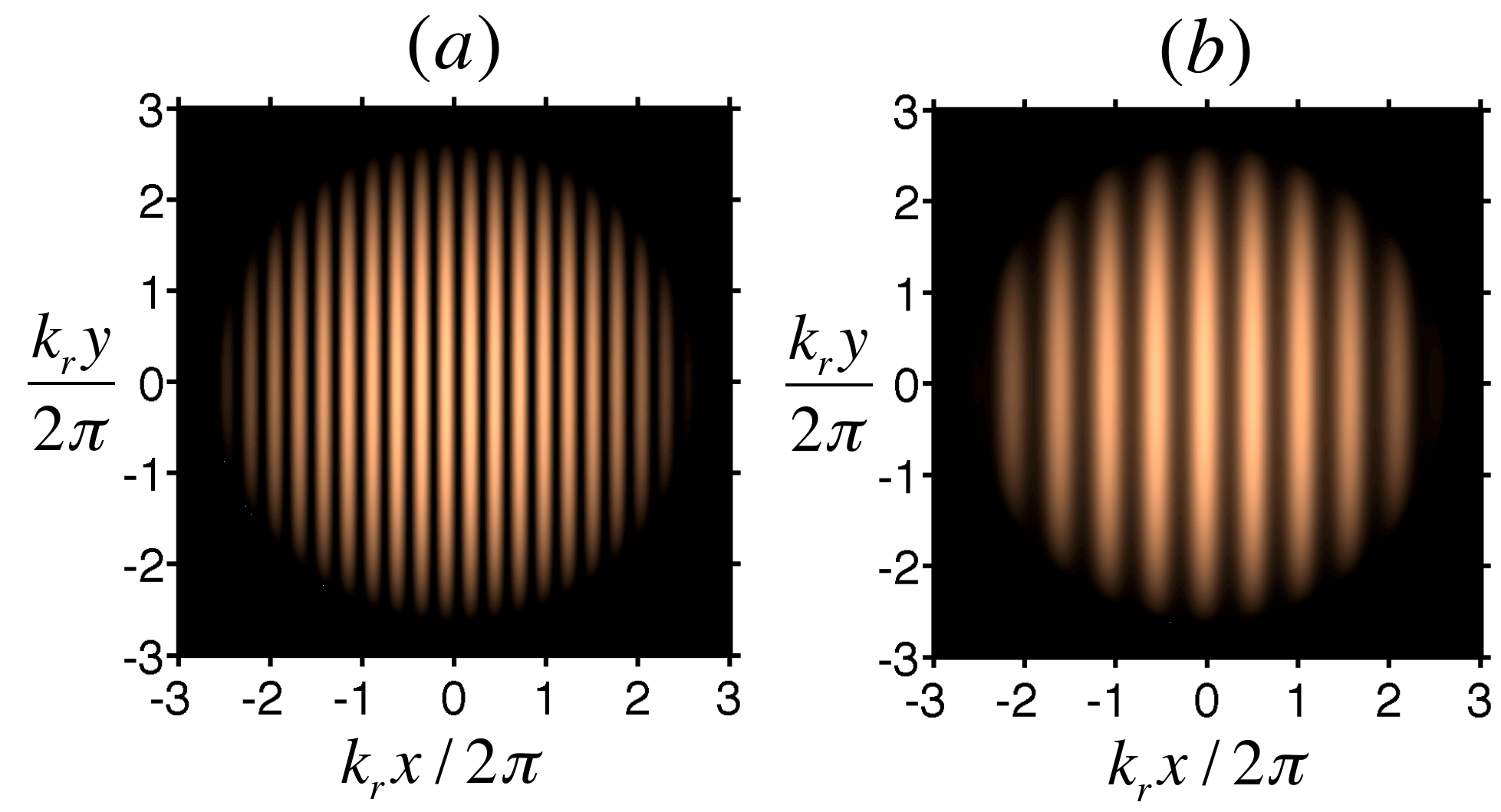}
\caption{ (color online) Two typical different kinds of stripe phases from the  three-minima band structure  in Fig.~\ref{regime}(c) for a BEC with $k_0=1.88 k_r$ when collisional interactions are present. Shown are the density distributions of the $m=0$ Zeeman state with (a) $g_0/g_2=1$ and   (b) $g_0/g_2$=-1, where $g_0>0$ for both cases (A harmonic trap is taken account for in the numerics by the local density approximation). The typical parameters realized experimentally as in \cite{NIST_n2} are $E_r/\hbar=1.1\times10^{4}{\rm Hz}$, $g_0=7.79\times 10^{-12} {\rm Hz} \, {\rm cm}^3$ and number of atoms $N=1.8\times 10^5$.}
\label{stripe}
\end{figure}
\section{Stripe phases from the three minima}
There are two possible phases for the BEC in the two minima regime when interactions are present, a plane wave phase and  a stripe phase depending on whether a single minimum or two minima are occupied~\cite{ho}.  For our three minima case, at the single particle level, the ground state for a BEC is triple degenerate and is described by $\Phi_m(x)=A_+\chi_m^{p_+}(x)+A_0\chi_m^{p_0}(x)+A_-\chi_m^{p_-}(x)$ where $A_{\pm,0}$ is the complex amplitude and  $\chi_m^{p_{\pm, 0}}(x)=e^{i p_{\pm, 0} x} \tilde {\chi}_m^{p_{\pm, 0}}$ with $ \tilde {\chi}_m^{p_{\pm, 0}}$ the spinor from the lowest eigenstate of the single particle Hamiltonian~(\ref{hamiltonian}) at the three minima of $k_x=\pm k_0, 0$.  At the many particle level,  the interaction will select which minimum or minima the system will condensate to by minimizing the interaction energy. For example, a single non-zero component of $A_{\pm, 0}$ means a plane wave phase, while two non-zero components of $A_{\pm, 0}$ create a standing wave phase ~\cite{zhai, ho}.  One interesting consequence of the triple-degenerate minima regime is that there are two different kinds of stripe phases with different wavelengths. When the BEC occupies the two minima at $k_x=\pm k_0$, the resulting stripe phase has twice smaller wavelength than that when the BEC occupies the two minima at
$k_x=k_0$ and $k_x=0$ or $k_x=-k_0$ and $k_x=0$.


The interaction Hamiltonian for a three component BEC is given by $\hat{\mathcal{H}}_{int}=\int d^3 r g_0 \hat{n}^2(r)+g_2\hat{\mathcal{F}}^2(r)$~\cite{ho_spinor, zhai},
where  $\hat{n}=\sum_m \hat{n}_m$ is the total population of the three Zeeman states, and $\hat{\mathcal{F}}=\phi_{\alpha}\mathcal{F}_{\alpha\beta}\phi_{\beta}$ is the spin-1 operator with $\mathcal{F}$ the spin-1 generalization of the Pauli matrix. The complex amplitudes $A_{\pm,0} $ are determined by minimizing the Gross-Pitaevski (GP) functional of the single particle Hamiltonian plus the interaction Hamiltonian. In the numerical investigations, apart from the plane wave phase as discussed before ~\cite{zhai, ho, li} which results from the occupation of a single minimum, we also find two kinds of stripe phases with different wavelengths. Fig.~\ref{stripe} shows a typical example  for the density of the $m=0$ Zeeman state at two different $g_0$ and $g_2$,  where the difference of factor two in wavelength is clearly seen. Note that since these two kinds of stripe phases have the same laser parameters, i.e., they stem from the same single particle dispersion,  the change of wavelength by the laser parameter as discussed in ~\cite{ho} can not explain their origin.

To better understand the nature of the two different stripe phases, we write the density of one of the Zeeman state (e.g., $m=0$) as $n_0=\phi_0 \phi_0^{*}$ with $\phi_0=(A_- e^{-ik_0x}a_-+A_0 a_0+A_+ e^{ik_0x}a_+)$, where $a_{\pm,0}$ (taken as real) is the $m=0$ component of the spinor wavefunction and $A_{\pm,0}$ the complex amplitude. Because of the nonzero overlap between the spinor part of the wavefunction, the density of each Zeeman state will develop a stripe structure. A straightforward calculation gives, $n_0=C+(C_1 e^{ik_0 x}+c.c)+(C_2e^{2ik_0x}+c.c)$, where $C=|A_-|^2|a_-|^2+|A_0|^2|a_0|^2+|A_+|^2|a_+|^2$, $C_1=a_0(A_-^*a_-A_0+A_+a_+A_0^*)$, and $C_2=A_-^*a_-A_+a_+$. For $g_0>0$, we find that when $g_0/g_2=1$, the values of $A_{\pm,0}$ which minimize the interaction energy, always have vanishing $A_0$, which means $C_1=0$, thus the wavelength of the stipe phase  is, $\pi/k_0$ in this case. For  $g_0/g_2=-1$,  the $A_{\pm,0}$ are all nonvanishing. But in this case, since $a_0$ is dominant (see for instance Fig.~\ref{rf} (b)), the term  $e^{ik_0x}$ is dominant over $e^{2ik_0x}$, which means the wavelength of the stripe in this case is given by $2\pi/k_0$. The two different stripe phases therefore originate from the separation between the three minima. A further physical insight may be gained by the fact that the weights of the three dressed branches are largely determined by the three Zeeman states themselves (e.g., see Fig.~\ref{rf}), i.e., the minimum at $\pm k_0$, $0$ by the Zeeman state $m=\mp 1,0$ respectively. So when $g_2>0  (g_2<0)$, the system wants a zero (large) spin to minimize (maximize) the interaction energy, consequently, $m=\pm 1 (m=1,0 $ or $m=-1,0)$ are occupied.  This shows the two different kinds of stripe phases have an interesting connection with the ferromagnetic and polar phases of spin-1 BEC \cite{ho_spinor} which is the true manifestation of spin-orbit coupling in this system, i.e., the structure in pseudo-spin space (ferromagnetic or polar) has been transferred to structures in orbit space (small or large wavelength stripe) since the three dressed spin states are represented by the three minima with different momentum.  It is certainly tempting to conjecture that other types of stripe phases will emerge when including more Zeeman states.  Since the two different kinds of stripe phases originate from the sign of $g_2$ when the $g_2$ term is dominating over the $g_0$ term (assumed positive), we could tune $g_0/g_2$ from $+1$ to $-1$ to see the transition of the wavelength from $\pi/k_0$ to $2\pi/k_0$. The dynamics of the transition would depend on the experimental details such as for instance non-adiabatic effects. 
The interaction parameters for observing these stripe phases can be reached experimentally by optical Feshbach resonance  \cite{feshbach_rmp} (see also recent experiment \cite{feshbach_nist} for Raman-induced Feshbach resonance in this setup), and the different stripe structures can be probed by Bragg light scattering~\cite{bragg} or detected by measuring the displacement of the atomic cloud after expansion when the trap is turned off~\cite{ho}.

\section{Conclusions and Discussions}
Motivated by the recent experiments on synthetic spin-orbit coupled quantum gases~\cite{NIST_n2, USTC_boson, ShanXi_fermi, MIT_fermi, quench_so} in the two minima regime, we investigated and showed how a three minima regime in this setup can be obtained. We found that when the contributions of the three Zeeman states are comparable, triple degenerate minima appear at the bottom of the band dispersions which can be translated into a spin-orbit coupled spin 1 quantum gas. We further found there are two different kinds of stripe phases in this setup which have their roots from the ferromagnetic and polar phases of spin-1 BEC, i.e., due to spin-orbit coupling, the structure in pseudo-spin space is manifested by the structure in orbit space.   The scenario can be generalised to create a spin-orbit coupled high spin quantum gas by including more Zeeman states. We note that recently a different experimental technique to create two minima in momentum space by shaking an optical lattice and {\it in situ} observation of ferromagnetic domains has been achieved \cite {shaking_chin}.  These techniques could also be applied to the three minima regime studied in this work.

\begin{acknowledgements}
We would like to thank Yu-Ju Lin and Ian Spielman for helpful comments and suggestions. Z.L.  acknowledges support from EPSRC grant No.  EP/I018514/1 and P. \"O. from EPSRC grant No. EP/J001392/1.
\end{acknowledgements}

\end{document}